\documentclass[twocolumn,pra,aps,reprint, superscriptaddress,showpacs,amsmath,amssymb,
aps, prl]{revtex4-1}
\usepackage{amssymb}
\usepackage{amsfonts}
\usepackage{amsmath}
\usepackage{cases}
\usepackage{stmaryrd}\usepackage{tipa}
\usepackage[dvips]{graphicx}
\usepackage{dcolumn}
\usepackage{bbold}
\usepackage{graphicx}
\usepackage{dcolumn}
\usepackage{bm}
\usepackage{color}
\usepackage{CJK}
\usepackage{subfigure}
 
\newcommand{\beq}{\begin{equation}}
\newcommand{\eeq}{\end{equation}}
\newcommand{\beqa}{\begin{eqnarray}}
\newcommand{\eeqa}{\end{eqnarray}}

\begin{document}

\title{Time and spatial parity operations with trapped ions}

\author{Xiao-Hang Cheng}
\affiliation{Department of Physics, Shanghai University, 200444 Shanghai, People's Republic of China}
\affiliation{Department of Physical Chemistry, University of the Basque Country UPV/EHU, Apartado 644, 48080 Bilbao, Spain}

\author{Unai Alvarez-Rodriguez}

\affiliation{Department of Physical Chemistry, University of the Basque Country UPV/EHU, Apartado 644, 48080 Bilbao, Spain}

\author{Lucas Lamata}

\affiliation{Department of Physical Chemistry, University of the Basque Country UPV/EHU, Apartado 644, 48080 Bilbao, Spain}

\author{Xi Chen}

\affiliation{Department of Physics, Shanghai University, 200444 Shanghai, People's Republic of China}

\author{Enrique Solano}

\affiliation{Department of Physical Chemistry, University of the Basque Country UPV/EHU, Apartado 644, 48080 Bilbao, Spain}
\affiliation{IKERBASQUE, Basque Foundation for Science, Maria Diaz de Haro 3, 48013 Bilbao, Spain}

\date{\today}

\begin{abstract}
We propose a physical implementation of time and spatial parity transformations, as well as Galilean boosts, in a trapped-ion quantum simulator. By embedding the simulated model into an enlarged simulating Hilbert space, these fundamental symmetry operations can be fully realized and measured with ion traps. We illustrate our proposal with analytical and numerical techniques of prototypical examples with state-of-the-art trapped-ion platforms. These results pave the way for the realization of time and spatial parity transformations in other models and quantum platforms.
\end{abstract}


\maketitle
\section{Introduction}
In the last decade, observing quantum phenomena that are difficult or even impossible to detect in the laboratory has been possible through the concept of quantum simulation \cite{qs}. Originally an idea of Richard Feynman \cite{Feynman}, it is based on implementing a complex quantum dynamics on a controllable quantum system. Many proposals and experiments on quantum simulations in different controllable platforms such as trapped ions \cite{trapped ion, trapped ion2, trapped ion3}, superconducting circuits \cite{superconducting1, superconducting2, superconducting3}, ultracold gases \cite{ultracold gas1, ultracold gas2}, quantum photonics systems \cite{photonics1, photonics2}, and optical lattices \cite{optical lattice1, optical lattices2}, have been performed and have led to a deeper understanding of a wide variety of phenomena.

Up to now, proposed models and experimental realizations of quantum simulations with trapped ions have been realized in spin models \cite{spin models1, spin models2, spin models3}, quantum field theories \cite{QFT}, quantum phase transitions \cite{quantum phase transition}, many-body systems \cite{fermion lattice, Ising model, Holstein model}, fermionic and bosonic interactions \cite{fermionic and bosonic models}, relativistic quantum physics, including Dirac equation \textit{Zitterbewegung},  \cite{qs dirac} and its realization in the laboratory \cite{nature}, Klein paradox \cite{Klein tunneling and Dirac potentials, Klein Paradox}, and interacting Dirac particles \cite{dirac particles}, among others. Recently, an implementation of the Majorana equation and unphysical quantum operations was proposed~\cite{majorana1}, and experimentally realized~\cite{SzameitMajo,KihwanMajo}. In addition, U. Alvarez-Rodriguez \textit{et al.} \cite{noncausal} have developed a mathematical formulation of an \textit{enlarged space} or \textit{embedding space} to perform linear transformations between space-time coordinates in a general quantum simulator. However, the implementation of these concepts in a trapped-ion simulator, including parity $\mathcal{P}$ operations, has not yet been analyzed.

In this Letter, we propose the realization of time and spatial parity operations, as well as Galilean boosts, in a trapped-ion quantum simulator. We perform analytical and numerical calculations in paradigmatic examples to illustrate our protocol, which is based on state-of-the-art trapped-ion technologies. We show that this proposal, including state initialization, dynamics, and measurement, can be efficiently implemented in current experiments. While Ref. \cite{noncausal} focuses on the underlying mathematical properties behind the theoretical protocol for Galilean transformations, here we propose a realistic implementation in trapped-ion systems. Moreover, we develop a toolbox for trapped ions that can be used to implement reference frame transformations for any given dynamical equations. New interesting types of simulations can emerge by adding the reference frame transformation to the toolbox of possible operations. This work significantly advances the field of quantum simulations of unphysical operations and establishes a path for implementing time and spatial parities in quantum optics systems. As a further scope, these results may allow us to enhance our capabilities when studying many-body interacting systems and their symmetries.

\begin{figure}[setup]
{\includegraphics[width=0.9 \linewidth]{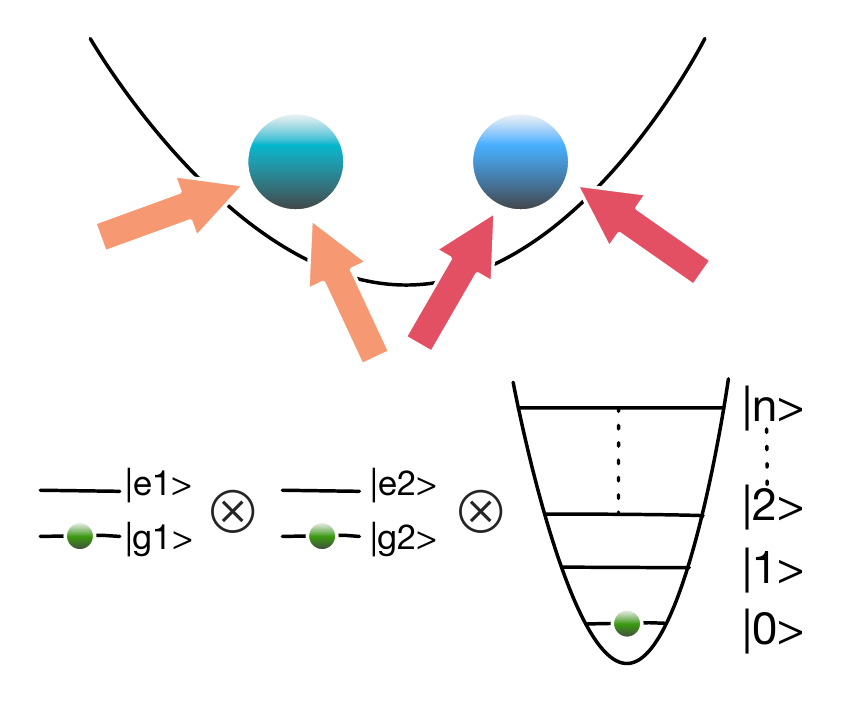}}
\caption{(color online) Scheme of the proposed experiment for the implementation of time, spatial, and Galilean transformations. Two ions are needed for the time parity and Galilean boost, while a single ion suffices for the spatial parity. \label{setup}}
\end{figure}

The formalism introduced in Ref.~\cite{noncausal} allows us to implement the quantum simulation of reference frame transformations in the lab. This symmetry transformation is described by a linear relation between the initial $(t,x)$ and the final $(t',x')$ coordinates, $x'_i=\sum_{ij} \alpha_{ij} x_j$, $i,j=0,1$. The spinor in the enlarged space is defined as $\Psi(x, t)=(\psi^e, \psi^o)^T$, where the even and odd part of any wave function can be expressed as $\psi^{e,o}=\frac{1}{2}[\psi(x, t)\pm\psi(x', t')]$. Therefore, the dynamical information of $\psi(x,t)$ and $\psi(x',t')$ is encoded in the evolution of $\Psi$. Moreover, through a judicious choice of measurement observables, one can perform a reference frame transformation via a local $\sigma^z$ operator, or even observe spacetime correlation functions between different reference frames. Throughout this paper we consider the evolution in the simulated Hilbert space as given by the Schr\"odinger equation $i\partial_t \psi=-ic\partial_x\psi$, where $c$ is a simulated speed of light and we fix $\hbar=1$. The corresponding equation for $\Psi$ in the embedding space, for arbitrary Galilean transformations, may be  written as
\begin{equation}
i \partial_{t} \Psi = -i  (\tilde{\alpha}_1 \mathbb{1} +\tilde{\alpha}_2 \sigma^x ) \partial_x \Psi, 
\end{equation}
where $\tilde{\alpha}_{1,2}=[c(\alpha_{11}\pm \alpha_{00}) \mp \alpha_{10}]/(2\alpha_{11})$.
We explain now how to use this representation for the implementation in a trapped-ion system. In the Lamb-Dicke regime, $\eta\sqrt{\langle(a+a^{\dag})^2\rangle}\ll1$, the Hamiltonian describing the interaction between an ion and a laser driving is \cite{trapped ion}
\begin{equation}
{\cal H}(t)=\Omega_0\sigma^+[1+i\eta(ae^{-i\nu t}+a^\dag e^{i\nu t})]e^{i(\phi-\delta t)}+{\rm H.c.},\nonumber
\end{equation}
where $\delta=\omega-\omega_0$ is the laser detuning, $\eta=k\sqrt{1/2m\nu}$ is the Lamb-Dicke parameter \cite{trapped ion}, $k$ is the wave number of the external field, $m$ is the mass of ion, $\nu$ is the frequency of a static potential harmonic oscillator, $\Omega_0$ is the coupling strength, $a$ and $a^{\dag}$ are the annihilation and creation operators of any suitable vibrational mode of the ion string, that we choose to be the center of mass motional mode, and $\phi$ is the field phase. In spin-1/2 language, $\sigma^+=|e\rangle\langle g|=(\sigma^x+i\sigma^y) / 2$, $\sigma^-=|g\rangle\langle e|=(\sigma^x-i\sigma^y) / 2$.

When $\delta=0$, a carrier resonance can be obtained with Hamiltonian $H_c=\Omega(\sigma^+e^{i\phi}+\sigma^-e^{-i\phi})$. A red-sideband, also known as Jaynes-Cummings (JC) interaction, is realised in the case of $\delta=-\nu$. This Hamiltonian is written as $H_r=\tilde{\Omega}\eta(a\sigma^+e^{i\phi_r}+a^\dag\sigma^-e^{-i\phi_r})$. Respectively, when $\delta=\nu$, a blue-sideband, anti-Jaynes-Cummings (AJC) interaction can be achieved, and its Hamiltonian can be expressed as $H_b=\tilde{\Omega}\eta(a^\dag\sigma^+e^{i\phi_b}+a\sigma^-e^{-i\phi_b})$. We illustrate now how to apply these techniques to generate the time and spatial parity transformations in trapped ions.

\section{Time parity transformation} 
As a first example, we show how to use two trapped ions to simulate a time parity transformation, $(t, x)\rightarrow(-t, x)$, $(\alpha_{00}, \alpha_{01}, \alpha_{10}, \alpha_{11})=(-1, 0, 0, 1)$. Here, we choose the Hamiltonian in the simulated space as a time-independent one, $H=H^e=c p$ and the momentum $p>0$, which describes a massless Dirac Hamiltonian without the internal degree of freedom. The corresponding one-dimensional Schr\"{o}dinger equation in the enlarged space can be expressed as
\begin{equation}
i \partial_t \Psi = \sigma_1^x c \hat{p} \Psi.
\end{equation}
The Hamiltonian in the enlarged space, $\mathcal{H}=\sigma_1^x\otimes H=\sigma_1^x c \hat{p}$, where $\sigma_1^x$ is the Pauli operator acting on ion 1, which can be realized by implementing a blue- and red-sideband simultaneously \cite{trapped ion, qs dirac},
\begin{equation}
\mathcal{H}=\eta\tilde{\Omega}(\sigma_1^+ a^\dag e^{i\phi_b}+\sigma_1^- a e^{-i\phi_b})+\eta\tilde{\Omega}(\sigma_1^+ a e^{i\phi_r}+\sigma_1^- a^\dag e^{-i\phi_r}),
\end{equation}
with proper phases for blue- and red-sideband $\phi_b=\pi/2$, $\phi_r=-\pi/2$.  Here, $\eta\tilde{\Omega}=\frac{c}{2\Delta}$ and $i(a^{\dag}-a)/2=\hat{p}\Delta$ with $\Delta=\sqrt{1/2m\nu}$.  We depict in Fig.~\ref{setup}, a scheme of the experimental setup with two ions interacting with lasers.

The initial state in the enlarged space is given as,
\begin{equation}
\Psi(x,t=0)=\left(\begin{array}{cc} 1 \\ 0 \end{array}\right)\otimes \psi(x,t=0),
\end{equation}
where $\psi(x,t=0)$ can be described as a Gaussian wave packet, $\psi(x,t=0)=\psi_0(x,t=0)e^{ip_0x}=(\sqrt{\sqrt{2\pi}\Delta})^{-1}e^{-x^2/4\Delta^2}e^{ip_0x}$. In a trapped-ion setup, this can be achieved by cooling the motional mode to the ground state, which is a Gaussian, and displacing it by simultaneous red and blue sidebands with the Hamiltonian $p_0 \hat{x} \sigma^x_2$, where $\hat{x}=\Delta (a+ a^\dag)$, with different phases in the $a$ and $a^\dag$ operators as compared to the simulating Hamiltonian $\mathcal{H}$. This will allow one to achieve an average $p_0$ momentum, using the auxiliary second ion initialized in an eigenstate of the $\sigma^x_2$ operator.
After applying the evolution propagator $\exp(-i\mathcal{H}t)$, we can evolve the state for any time. The solution reads
\begin{eqnarray}
&&\Psi(x,t)= \frac{1}{2\sqrt{\sqrt{2\pi}\Delta}}   \\ &&\times\left(\begin{array}{cc} e^{-\frac{(ct+x)^2}{4\Delta^2}}e^{ip_0(ct+x)}+e^{-\frac{(ct-x)^2}{4\Delta^2}}e^{-ip_0(ct-x)} \\ -e^{-\frac{(ct+x)^2}{4\Delta^2}}e^{ip_0(ct+x)}+e^{-\frac{(ct-x)^2}{4\Delta^2}}e^{-ip_0(ct-x)} \end{array}\right).\nonumber
\end{eqnarray}
Furthermore, the quantum states in the simulated spaces are obtained reversing the initial mapping,
\begin{eqnarray}
\psi(x,t)&=&(1,1)\Psi(x,t)=\frac{1}{\sqrt{\sqrt{2\pi}\Delta}}e^{-\frac{(ct-x)^2}{4\Delta^2}}e^{-ip_0(ct-x)},\nonumber\\
\psi(x,-t)&=&(1,1)\sigma^z\Psi(x,t)=\frac{1}{\sqrt{\sqrt{2\pi}\Delta}}e^{-\frac{(ct+x)^2}{4\Delta^2}}e^{ip_0(ct+x)}.\vspace{-0.2cm}\nonumber\\\label{wavepacketstime}
\end{eqnarray}

We plot in Fig.~\ref{WavepacketsFig}(a) the initial wavepacket in the simulated space, and in Fig.~\ref{WavepacketsFig}(b) the evolved and time-parity-transformed wavepackets in Eq.~(\ref{wavepacketstime}).
We calculate now the position average values in the simulated space for the different reference frames and their correlation,
\begin{equation}
\langle \hat{x} \rangle_{\psi(x,t)}=ct, \langle \hat{x} \rangle_{\psi(x,-t)}=-ct, \langle \hat{x} \rangle_{\psi(x,t),\psi(x,-t)}=0.
\end{equation}

The spacetime correlations may be measured in different ways with current ion technology. The one we introduce here extends the physical principle employed in \cite{nature}. The measurement is performed upon the $\sigma_1^z$ observable of the first ion, associated with the enlarged space degree of freedom, via fluorescence detection. To achieve this, a state-dependent displacement operator $U=\exp(-ik\hat{x}\sigma_1^x/2)$ is applied to the internal state of this ion and the joint mode, with $A=U^\dag \sigma_1^z U=\cos(k\hat{x})\sigma_1^z+\sin(k\hat{x})\sigma_1^y$~\cite{nature}.  In order to detect, e.g., the spacetime correlation $\langle \hat{x} \rangle_{\psi(x,t),\psi(x,-t)}$ in the simulated space, one should measure $\langle \hat{x} (\sigma_1^x +\mathbb{1}) \sigma_1^z \rangle_{\Psi(x,t)}$ following Eq.~(\ref{wavepacketstime}). Here, the operator $(\sigma_1^x+\mathbb{1}) \sigma_1^z$ acts on the qubit degree of freedom of the enlarged space, and the measurement can be decomposed into two parts, one for each summand, $\langle \hat{x} \sigma_1^z \rangle_{\Psi(x,t)}$, and $-i\langle \hat{x}\sigma_1^y \rangle_{\Psi(x,t)}$. These two measurements can be obtained from the derivative of the $A$ observable with respect to $k$ in the limit $k\langle \hat{x}\rangle\ll 1$, in which $\partial_k\langle A\rangle\approx\langle \hat{x}\sigma_1^y \rangle$, with a local rotation in the first case to change $\sigma_1^y$ into $\sigma_1^z$. Moreover, computing $\langle A\rangle$ for sizable $k$, for the cases of initial $\sigma_1^z$ and $\sigma_1^y$ eigenstates in the internal state, allows one to obtain $\cos(k\hat{x})$ and $\sin(k\hat{x})$. Via Fourier transform, we can obtain the position wavepacket probability distribution. The previous procedure enables us, among other things, to compute spacetime correlation functions without full tomography, which can reduce significantly the required resources.

\section{Spatial parity transformation}
The second case we consider is the simulation of a spatial parity transformation,  $(t, x)\rightarrow(t, -x)$, $(\alpha_{00}, \alpha_{01}, \alpha_{10}, \alpha_{11})=(1, 0, 0, -1)$. The initial state in the simulated space coincides with the one of time parity, $\psi(x, t=0)=(\sqrt{\sqrt{2\pi}\Delta})^{-1}e^{-x^2/4\Delta^2}e^{i p_0 x}$. So does the Hamiltonian in the simulated space, $H=H^e=c \hat{p}$. It is obvious that $\psi(-x, t=0)=(\sqrt{\sqrt{2\pi}\Delta})^{-1}e^{-x^2/4\Delta^2}e^{-i p_0 x}$. The Hamiltonian in the enlarged space $\mathcal{H}$ goes to $\sigma_1^x \otimes H^e$,
and the initial spinor in the enlarged space can be expressed as
\begin{equation}
\Psi(x,0)=\frac{1}{2\sqrt{\sqrt{2\pi}\Delta}}e^{-\frac{x^2}{4\Delta^2}}\left(\begin{array}{cc}e^{i p_0 x}+e^{-i p_0 x} \\ e^{i p_0 x}-e^{-i p_0 x}\end{array}\right).
\end{equation}

This can be achieved by initializing the internal state associated with the enlarged space in the $(1, 0)^T$ state, cooling the motional mode to the ground state, which is a Gaussian, and performing a conditional displacement of the motional state with the Hamiltonian $p_0 \hat{x}\sigma_1^x$. We point out that for spatial parity one ion suffices for both state initialization and simulation.

\begin{figure}[wavepackets]
\centering
{\includegraphics[width=1 \linewidth]{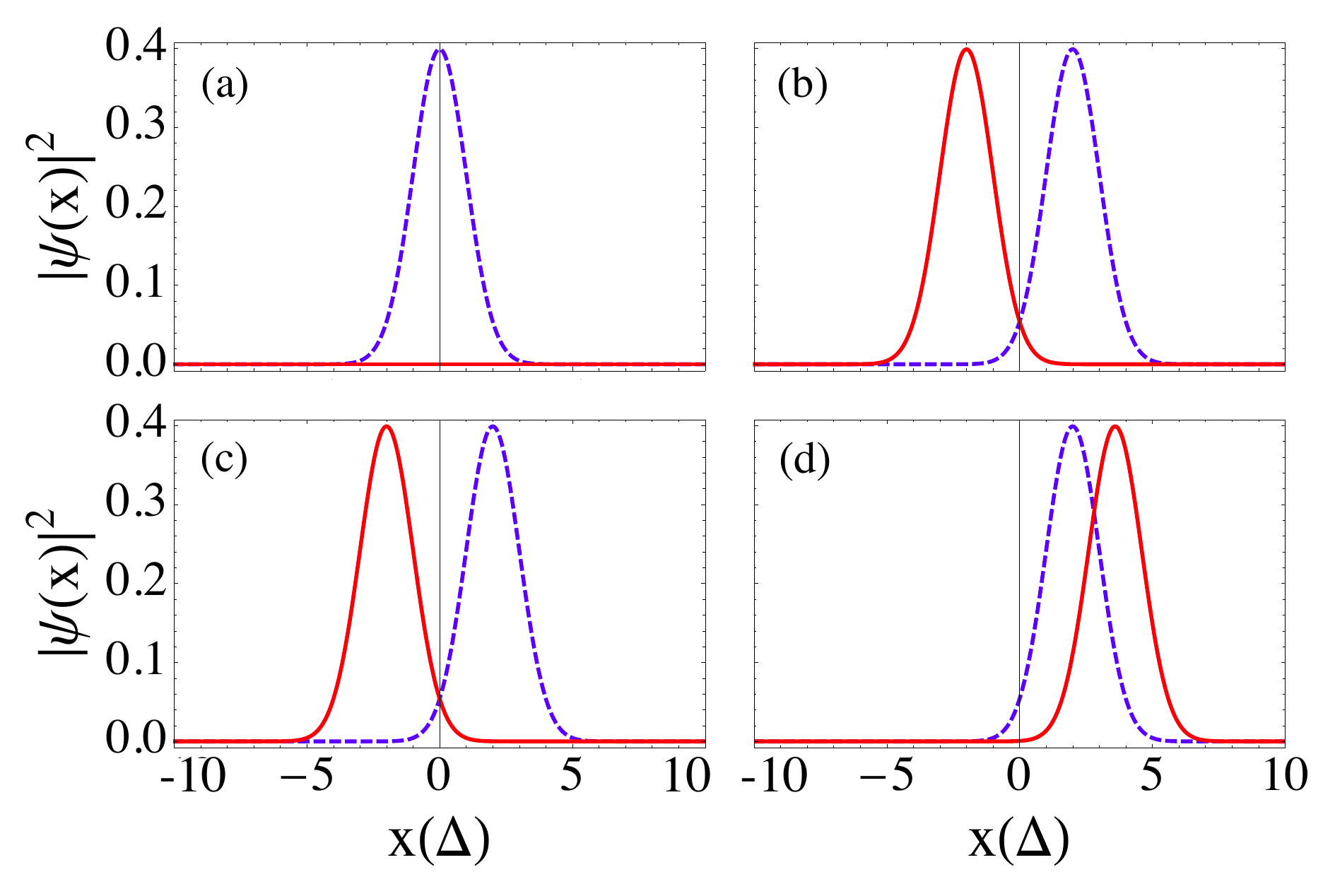}}
\caption{(color online) (a) Probability distribution $|\psi(x)|^2$ at time $t=0$. Probability distribution $|\psi(x)|^2$ at time $t=2\times10^3\Delta$  with $c=10^{-3}$ and $v/c = 0.8$ for time parity (b), spatial parity (c), and Galilean Boost (d). Dashed blue lines are time evolutions and solid red ones are the corresponding Galilean group transformations.\label{WavepacketsFig}}
\end{figure}

Accordingly, the state in the enlarged space considering time-evolution results,
\begin{eqnarray}
&&\Psi(x,t)=  \frac{1}{2\sqrt{\sqrt{2\pi}\Delta}}\\ &&\times\left(\begin{array}{cc} e^{-\frac{(ct-x)^2}{4\Delta^2}}e^{-ip_0(ct-x)}+e^{-\frac{(ct+x)^2}{4\Delta^2}}e^{-ip_0(ct+x)} \\ e^{-\frac{(ct-x)^2}{4\Delta^2}}e^{-ip_0(ct-x)}-e^{-\frac{(ct+x)^2}{4\Delta^2}}e^{-ip_0(ct+x)} \end{array}\right). \nonumber
\end{eqnarray}
As in the previous case, in order to recover the wavefunction in each of the simulated spaces we add or subtract the even and odd components of the spinor $\Psi$.
\begin{eqnarray}
\psi(x,t)&=&\frac{1}{\sqrt{\sqrt{2\pi}\Delta}} e^{\frac{-(ct-x)^2}{4\Delta^2}}e^{-ip_0(ct-x)},\label{SPp} \\
\psi(-x,t)&=&\frac{1}{\sqrt{\sqrt{2\pi}\Delta}} e^{\frac{-(ct+x)^2}{4\Delta^2}}e^{-ip_0(ct+x)}.\label{SPm}
\end{eqnarray}
We plot in Fig.~\ref{WavepacketsFig}(c) the evolved and spatial-parity-transformed wavepackets in Eqs.~(\ref{SPp}) and~(\ref{SPm}).
We also obtain the expectation values in each of the correlations.
\begin{eqnarray}
\langle \hat{x}\rangle_{\psi(x,t)}=ct, \qquad \langle \hat{x}\rangle_{\psi(-x,t)}=-ct, \nonumber
\end{eqnarray}
\begin{eqnarray}
&&\langle \hat{x}\rangle_{\psi(x,t),\psi(-x,t)}=\langle\Psi(x,t)|\left(\begin{array}{cc}1 & -1 \\ 1 & -1\end{array}\right)\hat{x}|\Psi(x,t)\rangle\nonumber\\&&=\langle\Psi(x,t)|\sigma_1^z\hat{x}|\Psi(x,t)\rangle-i\langle\Psi(x,t)|\sigma_1^y\hat{x}|\Psi(x,t)\rangle\nonumber\\&&=\langle\Psi(x,t)|\sigma_1^z\hat{x}|\Psi(x,t)\rangle-i\langle\Phi(x,t)|\sigma_1^z\hat{x}|\Phi(x,t)\rangle\nonumber\\&&=-2ip_0\Delta^2e^{-\frac{(ct)^2}{2\Delta^2}}e^{-2p_0^2\Delta^2} ,
\end{eqnarray}
where $\Phi(x,t)=e^{-i\pi \sigma_1^x/4}\Psi(x,t)$. These can be measured as in the previous example. We point out that the spatial parity case can be distinguished from the time parity case through the spacetime correlation, which is different in both cases. The computation of these correlations with our method makes full state tomography unnecessary.

\section{Galilean Boost}
Here, we propose the simulation of a reference frame change associated with a Galilean boost, $(t, x)\rightarrow(t, x-vt)$, $(\alpha_{00}, \alpha_{01}, \alpha_{10}, \alpha_{11})=(1, 0, -v, 1)$, where $v$ is the relative velocity~\cite{footnote}. Here, we consider the previous Hamiltonian and initial state in the simulated space. The corresponding Hamiltonian in the simulating enlarged space reads
\begin{equation}
\mathcal{H}(t)=\underbrace{\left(c+\frac{v}{2}\right)\mathbb{1}\hat{p}}_{\mathcal{H}_1(t)}\underbrace{-\frac{v}{2}\sigma^x{\hat p}}_{\mathcal{H}_2(t)}. \label{grill}
\end{equation}
Moreover, we can calculate the initial spinor state and compute the time evolution. The expression for the  quantum states in the simulated spaces reads,
\begin{eqnarray}
\psi(x,t)&=&\frac{1}{\sqrt{\sqrt{2\pi}\Delta}} e^{-\frac{(ct-x)^2}{4\Delta^2}}e^{-ip_0(ct-x)},\label{wavepacketsGalilean1}\\
\hspace{-0.3cm}\psi(x-vt,t)&=&\frac{1}{\sqrt{\sqrt{2\pi}\Delta}}  e^{-\frac{(ct-x+vt)^2}{4\Delta^2}}e^{-ip_0(ct-x+vt)},\label{wavepacketsGalilean2}
\end{eqnarray}

We plot in Fig.~\ref{WavepacketsFig}(d)  the evolution of the wavepackets with and without Galilean boost in Eqs.~(\ref{wavepacketsGalilean1}) and~(\ref{wavepacketsGalilean2}). Moreover, the expectation values for position $\hat{x}$ are
\begin{eqnarray}
&&\langle \hat{x}\rangle_{\psi(x,t)}=ct, \hspace{0.25cm} \langle \hat{x}\rangle_{\psi(x-vt,t)}=(c+v)t,\\
&&\langle \hat{x}\rangle_{\psi(x,t),\psi(x-vt,t)}=\frac{1}{2}t(2c+v)e^{-\frac{t^2v^2}{8\Delta^2}}e^{-ip_0tv}.
\end{eqnarray}

For the trapped-ion simulation the initialization of the spinor can be done similarly to the time parity case. For the subsequent dynamics we divide the Hamiltonian in Eq.~\eqref{grill} into two parts to implement its evolution in the trapped-ion system. To realize $\mathcal{H}_1$ in the laboratory, we propose to use a second auxiliary ion initialized in an eigenstate of $\sigma_2^x$,
\begin{equation}
\mathcal{H}_1|\Psi\rangle|+\rangle=\left(c+\frac{v}{2}\right)\mathbb{1}\hat{p}|\Psi\rangle|+\rangle\equiv\left(c+\frac{v}{2}\right)\sigma_2^x\hat{p}|\Psi\rangle|+\rangle.
\end{equation}
Then, the Hamiltonian can be implemented as
\begin{equation}
\label{H1Galilean}
\mathcal{H}_1'=\sigma_2^x\hat{p}\left(c+\frac{v}{2}\right)=i\eta\tilde{\Omega}_1\sigma_2^x(a^\dag-a),
\end{equation}
with $\eta\tilde{\Omega}_1=(c+\frac{v}{2})/2\Delta$.
Moreover, the second term in Eq.~\eqref{grill} can be realized as
\begin{equation}
\mathcal{H}_2=-\frac{v}{2}\sigma_1^x\hat{p}=i\eta\tilde{\Omega}_2\sigma_1^x(a^\dag-a),
\end{equation}
and with $\eta\tilde{\Omega}_2=-\frac{v}{4\Delta}$, through simultaneous red and blue sideband excitations upon the first ion.

\section{Discussion}
\begin{figure}[Cafidelities]
\centering
{\includegraphics[width=1 \linewidth]{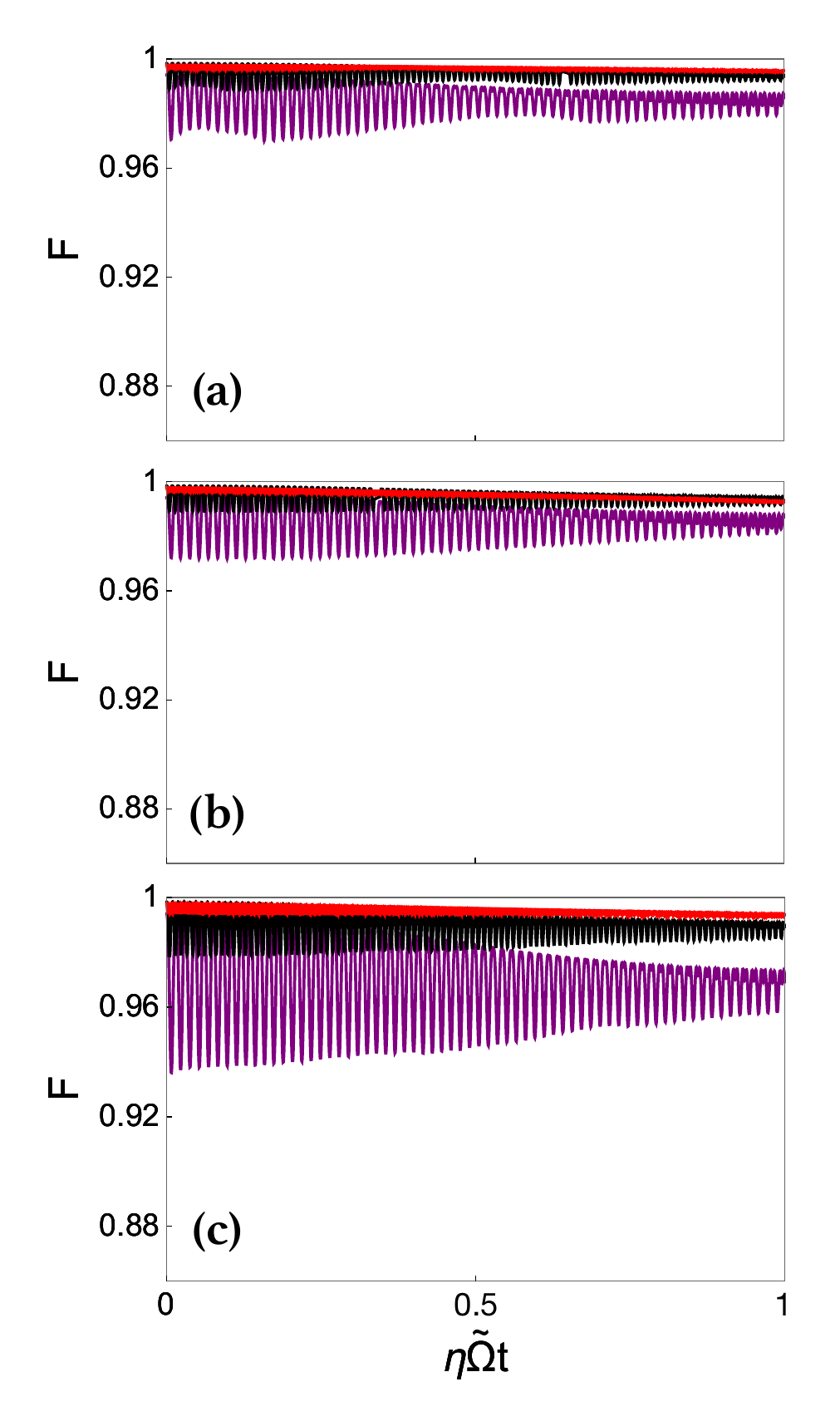}}
\caption{(color online) Fidelity $F={\rm Tr}[\rho|\psi_I\rangle\langle\psi_I|]$ of (a) time parity, (b) spatial parity, and (c) Galilean boost with $\tilde{\Omega}=0.01 \nu$ (red, upper), $\tilde{\Omega}=0.025 \nu$ (black, middle) and $\tilde{\Omega}=0.04 \nu$ (purple, lower). $\rho$ denotes the final state after initialization and dynamics, evolved with Eq.~(\ref{LabMasterEq}), and $|\psi_I\rangle$ denotes the ideal evolved state in absence of imperfections. We point out that $\tilde{\Omega}$ in the Galilean boost case (c) equals $\tilde{\Omega}_1$ in Eq.~(\ref{H1Galilean}). We consider the parameters of trapped ${\rm Ca}^{+}$ ion in some of the Innsbruck experiments~\cite{nature} $\eta=0.06$ and realistic decoherence rates~\cite{trapped ion2} $\Gamma_h=\Gamma_c=\Gamma_-=3.7\times10^{-7}\nu$ and $\Gamma_\phi=6.2\times10^{-7}\nu$.\label{Cafidelities}}
\end{figure}

\begin{figure}[Befidelities]
  \centering
{\includegraphics[width=1 \linewidth]{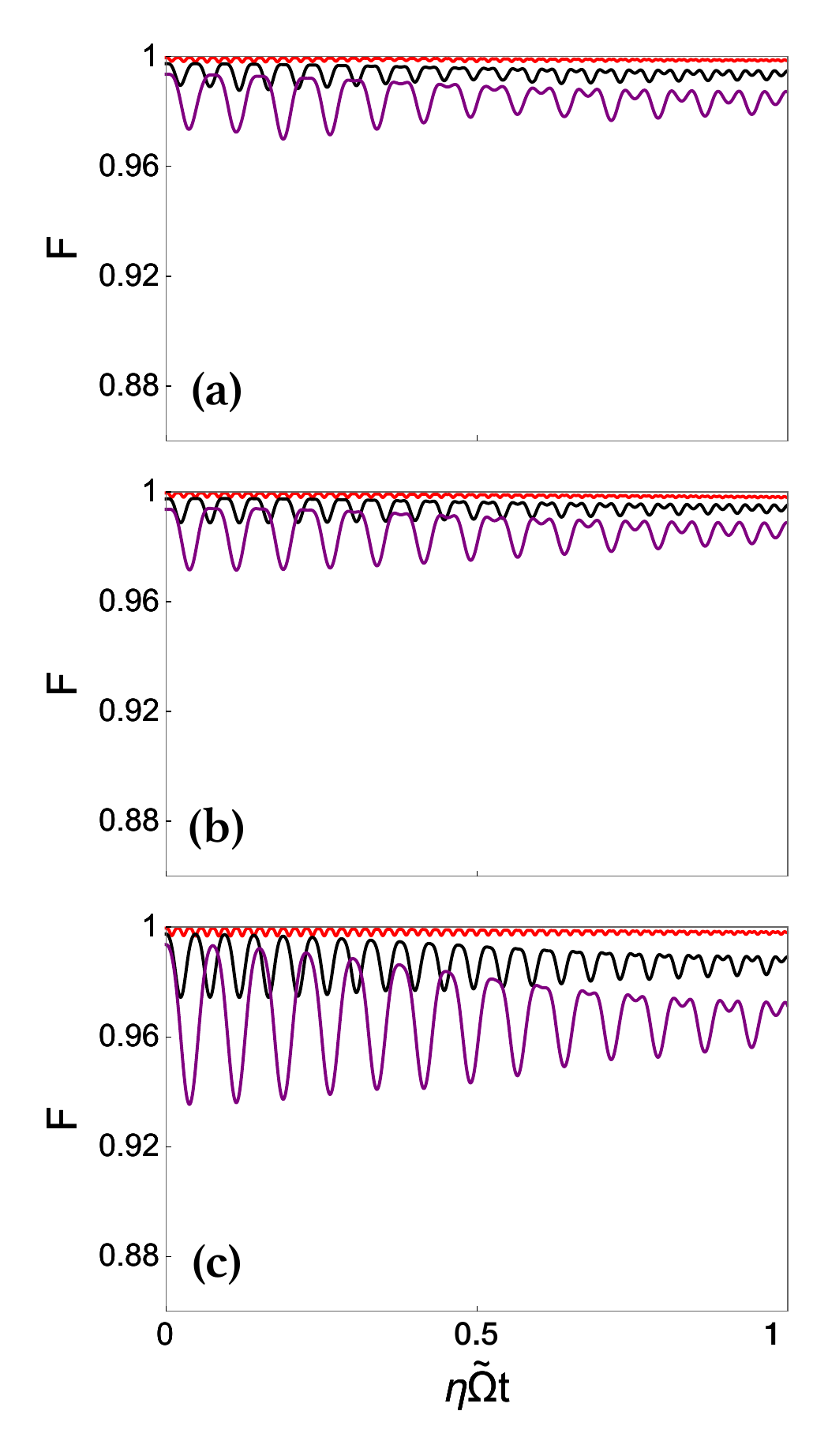}}
\caption{(color online) Fidelity of (a) time parity, (b) spatial parity, and (c) Galilean boost with $\tilde{\Omega}=0.01 \nu$ (red, upper), $\tilde{\Omega}=0.025 \nu$ (black, middle) and $\tilde{\Omega}=0.04 \nu$ (purple, lower),by considering trapped ${\rm Be}^{+}$ ion as an example. We use the parameters in some of the NIST experiments~\cite{trapped ion2}: Lamb-Dicke factor $\eta=0.3$, decoherence rates $\Gamma_h=\Gamma_c=\Gamma_-=3.7\times10^{-7}\nu$ and $\Gamma_\phi=6.2\times10^{-7}\nu$.\label{Befidelities}}
\end{figure}

To analyze the robustness of the simulating system, we computed the dynamics with a master equation including different decoherence sources. We considered unintended carrier transitions due to off-resonant coupling, heating $\Gamma_{h}$, phonon loss $\Gamma_{c}$, dephasing $\Gamma_\phi$, and spontaneous emission $\Gamma_{-}$,
\begin{eqnarray}\nonumber
\dot{\rho}=&&-i[{\cal H}_T,\rho]+\Gamma_{h} L(a^\dag)\rho+\Gamma_{c} L(a)\rho\\&&+\Gamma_\phi L(\sigma^z)\rho+\Gamma_{-}L(\sigma^{-})\rho,\label{LabMasterEq}
\end{eqnarray}
where the Lindblad superoperators are $L(\hat{X})\rho=(2\hat{X}\rho \hat{X}^{\dag}-\hat{X}^{\dag} \hat{X}\rho-\rho \hat{X}^{\dag} \hat{X})/2$. Here, ${\cal H}_T$ is the trapped-ion Hamiltonian corresponding to each of the three cases analyzed, namely, time parity, spatial parity, and Galilean boost. We include carrier and counterrotating sideband terms in the dynamics, i.e., without performing vibrational rotating-wave approximation. Therefore, this master equation accounts for all the significant decoherence and error sources present in current trapped-ion experiments.We plot in Figs.~\ref{Cafidelities}(a)-(c) and Figs.~\ref{Befidelities}(a)-(c) the fidelities of trapped calcium and beryllium ions after state initialization and evolution with the dynamics in Eq.~(\ref{LabMasterEq}) for the cases of time parity, spatial parity, and Galilean boost. For the state initialization part, we compute the dynamics with an equivalent master equation for the corrresponding initialization Hamiltonian as described for each case (a)-(c) in the text. We considered for the initialization time $p_0\Delta=\eta \tilde{\Omega} t=1$ in all cases. We point out that because of the relatively small atomic mass of beryllium, large trap frequency, sizable Lamb-Dicke factor and the corresponding large decoherence rates are introduced in the realistic experiments of NIST~\cite{trapped ion2}. Our work shows a significant feasibility in different trapped-ion setups.

\section{Conclusions}
To summarize, we have proposed the physical implementation of fundamental symmetry transformations including time and spatial parity with trapped ions. The formalism permits as well to perform Galilean boosts with the same technology. By embedding the simulated physical system into an enlarged Hilbert space living in the trapped-ion system, the proposed formalism can be carried out with current ion-trap setups. Furthermore, our work establishes a path for the realization of parity transformations in other quantum platforms and many-body interacting models.

\section{Acknowledgements}
We acknowledge funding from National Natural Science Foundation of China (61176118, 11474193), Shanghai Pujiang Program (13PJ1403000), Shuguang Program (14SG35), Program for Eastern Scholar, Specialized Research Fund for the Doctoral Program of Higher Education (2013310811003), Basque Government IT472-10 and BFI-2012-322, Spanish MINECO FIS2012-36673-C03-02, Ram\'{o}n y Cajal RYC-2012-11391, UPV/EHU EHUA14/04, UPV/EHU Grant No. UFI 272 11/55, PROMISCE, and SCALEQIT EU projects.

\end{document}